\documentclass[preprint,showpacs,prc]{revtex4}% Physical Review B
\usepackage{epsf}
\usepackage{amsmath,amsthm,amssymb}
\usepackage[usenames,dvipsnames]{xcolor}

\usepackage{dcolumn}% Align table columns on decimal point
\usepackage{bm}% bold math
\usepackage{graphicx,epsfig}

\graphicspath{{fig/}}
\everymath{\displaystyle}
\begin{document}           % End of preamble and beginning of text.

\title{General description of spin motion in storage rings in presence of
oscillating horizontal fields}

\author{Alexander J. Silenko} %\email{silenko@inp.minsk.by}
%\noaffiliation
\affiliation{Research Institute for Nuclear Problems, Belarusian State
University, Minsk 220030, Belarus} \affiliation{
Bogoliubov Laboratory of Theoretical Physics, Joint Institute for Nuclear Research, Dubna 141980, Russia}

\date{\today}

\begin {abstract}
The general theoretical description of the influence of oscillating horizontal magnetic and quasimagnetic fields on the spin evolution in storage rings is presented. Previous results are generalized to the case when both of the horizontal components of the oscillating field are nonzero and the vector of this field circumscribes an ellipse. General equations describing a behavior of all components of the polarization vector are derived and the case of an arbitrary initial polarization is considered.
The derivation is fulfilled in the case when the oscillation frequency is nonresonant.
The general spin evolution in storage rings conditioned by vertical betatron oscillations is calculated as an example.
\end{abstract}

%%%13.88.+e     Polarization in interactions and scattering
%%%29.27.Hj     Polarized beams
%%%%%%%\keywords{spin; polarized beams; storage rings}
\maketitle

%%%%%%%%%%%%%%%%%%%%%%%%%%%%%%%%%%%%%%%%%%%%%%%%%
\section{Introduction}
%%%%%%%%%%%%%%%%%%%%%%%%%%%%%%%%%%%%%%%%%%%%%%%%%

Particles and nuclei in accelerators and storage rings move in a main vertical magnetic field. Additionally (or alternatively), one can use a radial electric field in bending sections. Particle/nucleus beams are also governed by a radial magnetic or a vertical electric focusing field. All fields not only influence the motion of particles and nuclei but also act on their spins. In the present work, we confine ourselves to a description of spin effects. The spin rotation about the vertical direction is perturbed by either magnetic, $\bm B_\|$, or quasimagnetic, $(\bm v\times\bm E)_\|$, horizontal fields. The symbol $\|$ means a horizontal projection for any vector (Fig. 1).
Such fields are created by a focusing system and by rf devices placed into a storage ring. Due to an oscillatory motion of a particle or a nucleus in the storage ring \cite{Lee,MZ}, focusing magnetic and quasimagnetic fields acting on a particle spin also oscillate. A spin motion in storage rings in presence of resonant and nonresonant oscillatory horizontal fields has been calculated in Ref. \cite{EPJC2017}. The results obtained in that work do not, however, cover important cases when the two horizontal components of the oscillating field are nonzero. In particular, the spin motion at vertical betatron oscillations (BOs) belongs to such cases. When the phase difference between the two horizontal components is equal to $\pm\pi/2$, the vector $\bm B_\|$ (or $(\bm v\times\bm E)_\|$) circumscribes an ellipse.

In the present work, we calculate an evolution of the spin in the general case when the two horizontal components of the oscillating field are nonzero. We utilize the result obtained for the general description of the spin motion in presence of vertical BOs.

We use the system of units $\hbar=1,~c=1$ in some cumbersome formulas.

%%%%%%%%%%%%%%%%%%%%%%%%%%%%%%%%%%%%%%%%%%%%%%%%%
\section{Spin dynamics in constant vertical and oscillating horizontal fields}\label{VHField}
%%%%%%%%%%%%%%%%%%%%%%%%%%%%%%%%%%%%%%%%%%%%%%%%%

In this section, we use the approach elaborated in Ref. \cite{EPJC2017}.
Let a spinning particle or nucleus be placed into the constant vertical magnetic field $\bm B_0=B_0\bm e_z$ (see Fig. 1) or into the constant radial electric field and the angular velocity of the spin rotation in this field is equal to $\bm\omega_0=\omega_0\bm e_z$. Additional fields are conditioned by magnetic or electric focusing. The use of magnetic focusing brings a radial magnetic field and changes the vertical magnetic field ($B_z\neq B_0$). The particle spin is also affected by the oscillating horizontal magnetic or quasimagnetic field turning the spin about two horizontal axes. In this case, the spin-dependent part of the classical Hamiltonian is given by
\begin{equation}
\begin{array}{c}
H=\bm\omega_0\cdot\bm\zeta+[a_1\cos{(\omega t+\chi)}\bm e_x+a_2\sin{(\omega t+\chi)}\bm e_y]\cdot\bm\zeta,
%~~~ \bm\omega_0=-\frac
%{g_N\mu_N}{\hbar}\bm{B}_0,\\ \bm{\mathfrak{E}}=-\frac{g_N\mu_N}{2\hbar}\bm{\mathcal{B}},
\end{array}
\label{propt}\end{equation} where $\bm\zeta$ is the spin (pseudo)vector.
The quantities $\omega_0,\,\omega,\,a_1$, and $a_2$ can be positive and negative.

\begin{figure}[h]
\begin{center}
\includegraphics[width=7.5cm]{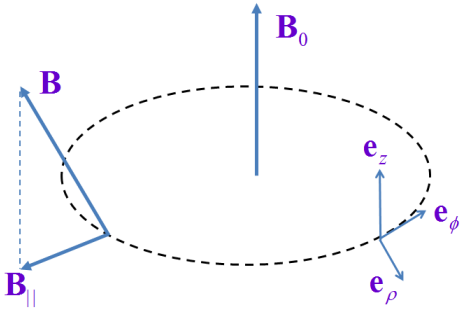} %%% {fig_color.eps for color)
%\includegraphics{fig1w}
%\vspace{1cm}
\caption{The storage ring geometry when the cylindrical coordinate system is used. The figure relates to the storage ring with the main magnetic field and magnetic focusing.}
\label{fig1}
\end{center}
\end{figure}

The angular velocity of the spin rotation is given by
\begin{equation}
\begin{array}{c}
\bm\Omega=\omega_0\bm e_z+a_1\cos{(\omega t+\chi)}\bm e_x+a_2\sin{(\omega t+\chi)}\bm e_y.
\end{array}
\label{anvelom} \end{equation}

The direction of the (pseudo)vector $\bm\omega_0$ defines the orientation of the so-called stable spin axis. In the absence of oscillating fields, the spin remains stable if it is initially aligned along this direction. If the initial spin orientation is different, the spin describes a cone around the direction of $\bm\omega_0$. The stable spin axis is a static quantity defined in the absence of oscillating fields.

%%%%%%% correction! %%%%%%%
Evidently, the horizontal field is the sum of two rotating fields whose amplitudes in the counterclockwise and clockwise directions are equal to $(a_1+a_2)/2$
%%%%%%%Evidently, the horizontal field is the sum of two rotating fields which amplitudes in the clockwise and counterclockwise directions are
%%%%%%%%%%%%%%%%%%%
and $(a_1-a_2)/2$, respectively. Resonance effects have been considered in detail in Ref. \cite{EPJC2017}. Therefore, we focus our attention on the nonresonance case. This case has been investigated in Ref. \cite{EPJC2017} only in the case of $a_2=0$.

Similarly to that work, we analyze the spin motion in the primed coordinate system rotating with the angular velocity $\bm\omega_0$:
\begin{equation}
\begin{array}{c}
\bm e'_x=\cos{(\omega_0 t)}\bm e_x+\sin{(\omega_0 t)}\bm e_y,\\ \bm e'_y=-\sin{(\omega_0 t)}\bm e_x+\cos{(\omega_0 t)}\bm e_y,\quad \bm e'_z=\bm e_z.
\end{array}
\label{contWvw} \end{equation}

If we denote $\bm{\mathfrak{K}}=\bm\Omega_\|=a_1\cos{(\omega t+\chi)}\bm e_x+a_2\sin{(\omega t+\chi)}\bm e_y$, the spin dynamics in the primed frame is defined by
\begin{equation}
\begin{array}{c}
\frac{d\bm\zeta'}{dt}=\bm{\mathfrak{K}}'\times\bm\zeta', \qquad \mathfrak{K}_x'= a_1\cos{(\omega t+\chi)}\cos{(\omega_0 t)}\\+a_2\sin{(\omega t+\chi)}\sin{(\omega_0 t)},\\
\mathfrak{K}_y'=-a_1\cos{(\omega t+\chi)}\sin{(\omega_0 t)}\\+a_2\sin{(\omega t+\chi)}\cos{(\omega_0 t)},\qquad \mathfrak{K}_z'=0.
\end{array}
\label{spinmon} \end{equation} Equivalently,
\begin{equation}
\begin{array}{c}
\mathfrak{K}_x'= \frac{a_1+a_2}{2}\cos{[(\omega_0-\omega) t-\chi]}\\+\frac{a_1-a_2}{2}\cos{[(\omega_0+\omega) t+\chi]},\\
\mathfrak{K}_y'=-\frac{a_1+a_2}{2}\sin{[(\omega_0-\omega) t-\chi]}\\-\frac{a_1-a_2}{2}\sin{[(\omega_0+\omega) t+\chi]},\qquad \mathfrak{K}_z'=0.
\end{array}
\label{spinmfn} \end{equation}

It is convenient to present the unit spin vector as a sum of two parts, $\bm\zeta(t)=\bm{\mathcal S}(t)+\bm\eta(t)$, where $\bm{\mathcal S}$ rotates with the angular velocity $\bm\omega_0$ \cite{EPJC2017}.
In this case, $\bm{\mathcal S}'$ is constant, $\bm{\mathcal S}'=\bm\zeta'(0)$, and $\bm\eta(0)=0$.

Equation (\ref{spinmfn}) shows that the periodical perturbation of the spin by the horizontal field is weak on condition that
\begin{equation}
\begin{array}{c}
|a_1+a_2|\ll|\omega_0-\omega|,\qquad |a_1-a_2|\ll|\omega_0+\omega|.
\end{array}
\label{cnd} \end{equation}
We suppose that this condition is valid. In this case, $|\bm\eta|\ll|\bm{\mathcal S}'|$ and the following equation can be used \cite{EPJC2017}:
\begin{equation}
\begin{array}{c}
\frac{d\bm\zeta'}{dt}=\frac{d\bm\eta'}{dt}=\bm{\mathfrak{K}}'\times\bm{\mathcal S}'.
\end{array}
\label{spinmno} \end{equation}

In the general case, the initial spin direction is defined by the spherical angles $\theta$ and $\psi$:
\begin{equation}
P_x(0)\!=\!\sin{\theta}\cos{\psi},\quad P_y(0)\!=\!\sin{\theta}\sin{\psi},\quad P_z(0)\!=\!\cos{\theta}, \label{pfndppfndp} \end{equation}
where $\bm P=\bm\zeta/s$ is the polarization vector and $s$ is the spin
quantum number. Dynamics of the vector $\bm{\mathcal S}$ is given by
\begin{equation}
\begin{array}{c}
{\mathcal S}_x(t)=s\sin{\theta}\cos{(\omega_0 t+\psi)},\\ {\mathcal S}_y(t)=s\sin{\theta}\sin{(\omega_0 t+\psi)},\qquad {\mathcal S}_z(t)=s\cos{\theta}.
\end{array}
\label{mtpk} \end{equation}

An integration on time results in the following evolution of the polarization vector:
\begin{widetext}
\begin{equation}
\begin{array}{c}
P_x(t)=\sin{\theta}\cos{(\omega_0 t+\psi)}+\frac12\Biggl(\frac{a_1-a_2}{\omega_0 +\omega}\biggl\{\cos{(\omega t+\chi)}\left[1-\cos{(\omega_0 +\omega)t}\right]-\sin{(\omega t+\chi)}\sin{(\omega_0 +\omega)t}\biggr\}\\+\frac{a_1+a_2}{\omega_0 -\omega}\biggl\{\cos{(\omega t+\chi)}\left[1-\cos{(\omega_0 -\omega)t}\right]+\sin{(\omega t+\chi)}\sin{(\omega_0 -\omega)t}\biggr\}\Biggr)\cos{\theta},\\
P_y(t)=\sin{\theta}\sin{(\omega_0 t+\psi)}+\frac12\Biggl(-\frac{a_1-a_2}{\omega_0 +\omega}\biggl\{\sin{(\omega t+\chi)}\left[1-\cos{(\omega_0 +\omega)t}\right]+\cos{(\omega t+\chi)}\sin{(\omega_0 +\omega)t}\biggr\}\\+\frac{a_1+a_2}{\omega_0 -\omega}\biggl\{\sin{(\omega t+\chi)}\left[1-\cos{(\omega_0 -\omega)t}\right]-\cos{(\omega t+\chi)}\sin{(\omega_0 -\omega)t}\biggr\}\Biggr)\cos{\theta},\\
P_z(t)=\cos{\theta}+\frac12\Biggl(\frac{a_1-a_2}{\omega_0 +\omega}\biggl\{\left[1-\cos{(\omega_0 +\omega)t}\right]\cos{(\psi+\chi)}+\sin{(\omega_0 +\omega)t}\sin{(\psi+\chi)}\biggr\}\\ +\frac{a_1+a_2}{\omega_0 -\omega}\biggl\{\left[1-\cos{(\omega_0 -\omega)t}\right]\cos{(\psi-\chi)}+\sin{(\omega_0 -\omega)t}\sin{(\psi-\chi)}\biggr\}\Biggr)\sin{\theta}.
\end{array}
\label{polvmon} \end{equation} %\end{widetext}

%\begin{widetext}
Equation (\ref{polvmon}) can be reduced to the form
\begin{equation}
\begin{array}{c}
P_x(t)=\sin{\theta}\cos{(\omega_0 t+\psi)}+\frac12\Biggl\{A_1\biggl[\cos{(\omega t +\chi)}-\cos{(\omega_0t -\chi)}\biggr]+A_2\biggl[\cos{(\omega t+\chi)}-\cos{(\omega_0t+\chi)}\biggr]\Biggr\}\cos{\theta},\\
P_y(t)=\sin{\theta}\sin{(\omega_0 t+\psi)}+\frac12\Biggl\{-A_1\biggl[\sin{(\omega t+\chi)}+\sin{(\omega_0t-\chi)}\biggr]
+A_2\biggl[\sin{(\omega t+\chi)}-\sin{(\omega_0t+\chi)}\biggr]\Biggr\}\cos{\theta},\\
P_z(t)=\cos{\theta}+\frac12\Biggl(A_1\biggl\{\cos{(\psi+\chi)}-\cos{[(\omega_0 +\omega)t+\psi+\chi]}\biggr\} %\\
+A_2\biggl\{\cos{(\psi-\chi)}-\cos{[(\omega_0 -\omega)t+\psi-\chi]}\biggr\}\Biggr)\sin{\theta},\\
A_1=\frac{a_1-a_2}{\omega_0 +\omega},\qquad A_2=\frac{a_1+a_2}{\omega_0 -\omega}.
\end{array}\label{polvmol} \end{equation}\end{widetext}

Equations (\ref{polvmon}) and (\ref{polvmol}) give the general description of the spin motion in storage rings in the presence of
oscillating horizontal fields. These equations agree with the corresponding equations obtained in Ref. \cite{EPJC2017} and their validity %of these equations
can be confirmed by calculating of the time derivative $d\bm P/(dt)$ and checking its consistency with Eq. (\ref{anvelom}).

%%%%%%%%%%%%%%%%%%%%%%%%%%%%%%%%%%%%%%%%%%%%%%%%%
\section{Influence of vertical betatron oscillations on spin evolution in storage rings}\label{verticalBO}
%%%%%%%%%%%%%%%%%%%%%%%%%%%%%%%%%%%%%%%%%%%%%%%%%

As an example of an application of the results obtained, we consider the spin evolution in storage rings affected by vertical BOs also called pitch oscillations. To simplify a needed derivation, we do not take into account an influence of radial betatron oscillations on spin dynamics considered in Ref. \cite{Fukuyama}.

It is important that the general equations defining spin dynamics can be successfully used not only for the Cartesian coordinates but also for the cylindrical and Frenet-Serret coordinates characterizing the spin turn relative to the momentum direction. In these cases, the external fields should also be expressed in the same coordinates. When the cylindrical coordinate system is used, one characterizes a position of a moving particle by the azimuthal angle $\phi$. Let the angular velocity of the particle revolution be equal to $\bm\omega$. The connection between the angular velocities of the spin rotation in the cylindrical and Cartesian coordinates is expressed by the equation \cite{RPJSTAB,JINRLettCylr} $\bm\Omega^{(cyl)}=\bm\Omega^{(Car)}-\omega_z\bm e_z$ $(\omega_z=\dot{\phi})$. The corresponding connection between the angular velocities of the spin rotation in the Frenet-Serret and Cartesian coordinates is given by (see, e.g., Refs. \cite{JINRLettCylr,FukuyamaSilenko}) $\bm\Omega^{(FS)}=\bm\Omega^{(Car)}-\bm\omega$.

In Sec. \ref{VHField}, we used the standard approach when the magnetic resonance (including nonresonance frequencies) is caused by an oscillating magnetic field orthogonal to a constant magnetic field and these fields are defined \emph{in the particle rest frame}. When the Thomas precession is taken into account and/or the particle possesses an electric dipole moment (EDM), an electric field in the particle rest frame can also be important (see Ref. \cite{PhysScr}). However, the Thomas-Bargmann-Michel-Telegdi equation \cite{Thomas-BMT} and its extension involving particles with the EDM (see Refs. \cite{FukuyamaSilenko,PhysScr} and references therein) allow one to describe the quasimagnetic resonance for \emph{moving particles}. In this case, all fields are defined in the lab frame while an existence of the quasimagnetic resonance is checked in the particle rest frame \cite{EPJC2017,PhysRevSTAB2017}. It is important that the rest frame fields can oscillate even if the lab frame fields are constant. This situation takes place due to the BOs of the beam. All the lab frame fields entering equations of motion are determined in a point defining a particle position in a given moment of time. As a result, the BOs make these fields to be oscillatory.

In the present study, we consider both electric and magnetic
focusing. Due to a cyclic motion of particles and nuclei in storage rings, it is more convenient to use the cylindrical coordinates instead of the Cartesian ones.
The angular velocity of the spin rotation in the cylindrical coordinate system is equal to \cite{RPJSTAB}
\begin{equation}
\begin{array}{c}
\bm\Omega^{(cyl)}=-\frac{e}{m}\left\{G\bm B-
\frac{G\gamma}{\gamma+1}\bm\beta(\bm\beta\cdot\bm B)\right.\\
+\left(\frac{1}{\gamma^2-1}-G\right)\left(\bm\beta\times\bm
E\right)+\frac{1}{\gamma}\left[\bm B_\|
-\frac{1}{\beta^2}\left(\bm\beta\times\bm
E\right)_\|\right]\\ \left.+ \frac{\eta}{2}\left(\bm
E-\frac{\gamma}{\gamma+1}\bm\beta(\bm\beta\cdot\bm
E)+\bm\beta\times\bm B\right)\right\}, % +o\bm e_z, ~~~
\end{array}\label{eqf}\end{equation} where $\bm\beta=\bm v/c,\,G=(g-2)/2,\,\eta=2mcd/(es),\,\gamma$ is the Lorentz factor, and $d$ is the EDM.
The vertical and horizontal components of the magnetic and quasimagnetic fields, $\bm B$ and $\bm\beta\times\bm E$, enter into Eq. (\ref{eqf}) with different factors.
In accelerator physics, the Frenet-Serret coordinate system is traditionally used \cite{Lee,LeeAP}.
%A comparison between the equations of the spin motion in the cylindrical and Frenet-Serret coordinate systems has been made in Ref. \cite{JINRLettCylr}.

When the radial BOs are not taken into account, the momentum vector lies in the plane formed by the vectors $\bm e_\phi$ and $\bm e_z$ and makes
the small angle $\vartheta=p_z/p$ with the axis $\bm e_\phi$. When $\vartheta=0$,
the spin rotates about the vertical axis with the angular frequency $\omega_0$. A focusing field leads to the oscillation of the angle $\vartheta$ with the pitch frequency $\omega$:
$$\vartheta=\vartheta_0\sin{(\omega t+\chi)}.$$

The equation of the vertical BO in the focusing electric field is
$$\frac{dp_z}{dt}=eE_z. $$
Since $p=m\beta\gamma~(c$ is omitted), \begin{equation}
eE_z=m\beta\gamma\omega\vartheta_0\cos{(\omega t+\chi)}.
\label{eq10}\end{equation}

In the focusing magnetic field, this equation takes the form
$$\frac{dp_z}{dt}=-e\lambda\beta B_\rho $$
or \begin{equation} eB_\rho=-\lambda
m\gamma\omega\vartheta_0\cos{(\omega t+\chi)},
\label{eq11}\end{equation} where $\lambda$ is equal to 1 and $-1$ for a
particle with a negative charge moving counterclockwise %($\mu^-$)
and for a particle with a positive charge moving clockwise, %($\mu^+$),
respectively.

If electric focusing is used,
\begin{equation}\begin{array}{c} -\frac emG\bm B=-\frac emGB_z\bm e_z=\omega_0\bm e_z, \\
\frac em\cdot\frac{G\gamma}{\gamma+1}\bm\beta(\bm\beta\cdot\bm B)=
\frac em\cdot\frac{G\gamma}{\gamma+1}\left(\beta_z^2B_z\bm
e_z+\lambda\beta\beta_zB_z\bm e_\phi\right)\\
=-\omega_0\frac{\gamma-1}{\gamma}\left[\vartheta_0^2\sin^2{(\omega t+\chi)}\bm
e_z+\lambda\vartheta_0\sin{(\omega t+\chi)}\bm e_\phi\right],\\
\frac em\left(G+\frac{1}{\gamma+1}\right)\left(\bm\beta\times\bm
E\right)_\|=\lambda\frac em\left(G+\frac{1}{\gamma+1}\right)\beta
E_z\bm e_\rho \\ =\lambda f\omega\vartheta_0\cos{(\omega t+\chi)}\bm e_\rho,
\end{array}\label{dded}\end{equation}
where $f$ is given by
\begin{equation} f=1+G\gamma-\frac{1+G}{\gamma}=1+G\beta^2\gamma-\frac{1}{\gamma}.
\label{eq7}\end{equation}

The angular velocity of the spin rotation is equal to
\begin{equation}\begin{array}{c}
\bm\Omega=\omega_0\left[1-\frac{\gamma-1}{\gamma}\vartheta_0^2\sin^2{(\omega t+\chi)}\right]\bm
e_z \\ -\lambda\frac{\gamma-1}{\gamma}\omega_0\vartheta_0\sin{(\omega t+\chi)}\bm
e_\phi \\ +\lambda f\omega\vartheta_0\cos{(\omega t+\chi)}\bm
e_\rho.
\end{array}\label{eq12}\end{equation}

If magnetic focusing is used,
$$\begin{array}{c} \bm B=B_z\bm e_z+B_\rho\bm e_\rho,\qquad
-\frac emGB_z\bm e_z=\omega_0\bm e_z, \\
\frac em\cdot\frac{G\gamma}{\gamma+1}\bm\beta(\bm\beta\cdot\bm B)=
\frac em\cdot\frac{G\gamma}{\gamma+1}\left(\beta_z^2B_z\bm
e_z \right.\\ \left.+\lambda\beta\beta_zB_z\bm e_\phi\right)
=-\frac{\gamma-1}{\gamma}\omega_0\left[\vartheta_0^2\sin^2{(\omega t+\chi)}\bm
e_z \right.\\ \left.+\lambda\vartheta_0\sin{(\omega t+\chi)}\bm e_\phi\right],\\
-\frac em\left(G+\frac{1}{\gamma}\right)\bm B_\|=-\frac
{e}{m\gamma}fB_\rho\bm e_\rho \\ =\lambda
f\omega\vartheta_0\cos{(\omega t+\chi)}\bm e_\rho,
\end{array}$$
where $f$ is defined by the different equation:
\begin{equation} f=1+G\gamma.
\label{eq8}\end{equation} The term proportional to $\bm\beta\cdot\bm B^{(foc)}$, where $\bm B^{(foc)}$ is the focusing magnetic field, can be neglected.

The angular velocity of the spin rotation is given by Eq. (\ref{eq12}), where $f$
is defined by Eq. (\ref{eq8}).

Taking into account radial BOs should bring some additional terms.

In Eq. (\ref{eq12}),  $\sin^2{(\omega
t+\chi)}=\bigl(1-\cos{[2(\omega t+\chi)]}\bigr)/2$. The small term oscillating with the angular frequency $2\omega$ can be neglected. As a result, Eq. (\ref{eq12}) takes the following final form:
\begin{equation}\begin{array}{c}
\bm\Omega=\omega_0\left(1-\frac{\gamma-1}{2\gamma}\vartheta_0^2\right)\bm
e_z-\lambda\frac{\gamma-1}{\gamma}\omega_0\vartheta_0\sin{(\omega t+\chi)}\bm
e_\phi\\ +\lambda f\omega\vartheta_0\cos{(\omega t+\chi)}\bm
e_\rho.
\end{array}\label{eqfinal}\end{equation} We should mention that the angular frequency of the spin rotation caused by the vertical BOs has not only the transversal components but also the longitudinal one.

Equations (\ref{eq12}) -- (\ref{eqfinal}) relate to the case of the main magnetic field. However, these equations can also be used when the bending radial electric field is applied together with the vertical magnetic field or instead of it. In this case, the main term in the equation of the spin motion takes the form
\begin{equation}\begin{array}{c} -\frac{e}{m}\left[G\bm B
+\left(\frac{1}{\gamma^2-1}-G\right)\left(\bm\beta\times\bm
E\right)\right]=-\frac{e}{m}\Bigl[GB_z\\
-\left(\frac{1}{\gamma^2-1}-G\right)\lambda\beta E_\rho\Bigr]\bm e_z=\omega_0\bm e_z.
\end{array}\label{maintrm}\end{equation}
The angular velocity of the spin rotation is given by
\begin{equation}\begin{array}{c}
\bm\Omega=\left(\omega_0-\omega'\frac{\gamma\!-\!1}{2\gamma}\vartheta_0^2\right)\bm
e_z-\lambda\frac{\gamma\!-\!1}{\gamma}\omega'\vartheta_0\sin{(\omega t+\chi)}\bm
e_\phi\\ +\lambda f\omega\vartheta_0\cos{(\omega t+\chi)}\bm
e_\rho,\qquad \omega'=-\frac{e}{m}GB_z,
\end{array}\label{eqfinll}\end{equation} where $B_z$
is the magnetic field averaged over the ring circumference. When one uses only the electric field in bending sections, $B_z=0$ and $\omega'=0$. The quantity $\omega_0$ is expressed by Eq. (\ref{maintrm}).

Since Eq. (\ref{eqfinll}) is a special case of Eq. (\ref{anvelom}), the influence of the vertical BOs on the spin evolution in storage rings is exhaustively described by the general equations (\ref{polvmon}) and (\ref{polvmol}). These equations also define spin dynamics affected by rf devices with fields circumscribing ellipses.

%%%%%%%%%%%%%%%%%%%%%%%%%%%%%%%%%%%%%%%%%%%%%%%%%
\section{Summary}\label{summary}
%%%%%%%%%%%%%%%%%%%%%%%%%%%%%%%%%%%%%%%%%%%%%%%%%

In the present paper, the general theoretical description of the influence of oscillating horizontal magnetic and quasimagnetic fields on the spin evolution in storage rings has been given. Unlike Ref. \cite{EPJC2017}, we have supposed that both of the horizontal components of the oscillating field are nonzero and the vector of this field circumscribes an ellipse. We have derived the general equations describing a
behavior of all components of the polarization vector and have considered the case of an arbitrary initial polarization.
The derivation has been fulfilled in the case when the condition (\ref{cnd}) is satisfied and the oscillation frequency is nonresonant.
The general spin evolution in storage rings conditioned by the vertical BOs has been calculated as an example. A precise calculation of a contribution of the BOs is necessary for measurements of a spin tune carrying out in the framework of the storage-ring EDM experiments \cite{PhysRevSTAB2017,PhysRevLett2017}.

%%%%%%%%%%%%%%%%%%%%%%%%%%%%%%%%%%%%%%%%%%%%%%%%%
\section*{Acknowledgements}
%%%%%%%%%%%%%%%%%%%%%%%%%%%%%%%%%%%%%%%%%%%%%%%%%

The author is grateful to T. Fukuyama and N. Nikolaev for useful discussions and acknowledges the support by the Belarusian Republican Foundation for Fundamental Research
(Grant No. $\Phi$16D-004) and
by the Heisenberg-Landau program of the German Ministry for
Science and Technology (Bundesministerium f\"{u}r Bildung und
Forschung).

\end{document}